\documentstyle[12pt,aaspp4,epsf]{article}
\begin{document}

\title{NON-THERMAL EMISSION FROM RELATIVISTIC ELECTRONS IN CLUSTERS OF 
       GALAXIES: A MERGER SHOCK ACCELERATION MODEL}

\author{Motokazu Takizawa}
\affil{Department of Astronomy, Faculty of Science, Kyoto University, 
       Sakyo-ku, Kyoto 606-8502, Japan}
\affil{Research Center for the Early Universe, Graduate School of Science,
       University of Tokyo, Bunkyo-ku, Tokyo, 113-0033, Japan}
\authoremail{takizawa@astron.s.u-tokyo.ac.jp}

\and

\author{Tsuguya Naito}
\affil{Yamanashi Gakuin University, Department of Management Information,
  Sakaori 2-4-5, Koufu-si, Yamanashi-ken, 400-8575, Japan}
\authoremail{tsuguya@ygu.ac.jp}

\begin{abstract}
We have investigated evolution of non-thermal emission from relativistic 
electrons accelerated at around the shock fronts during merger of 
clusters of galaxies. We estimate synchrotron radio emission and
inverse Compton scattering of cosmic microwave background photons 
from extreme ultraviolet (EUV) to hard X-ray range.
The hard X-ray emission is most luminous in the later 
stage of merger. Both hard X-ray and radio emissions
are luminous only while signatures of merging events are clearly seen
in thermal intracluster medium (ICM). On the other hand, EUV radiation
is still luminous after the system has relaxed. Propagation of shock waves and 
bulk-flow motion of ICM play crucial roles to extend radio halos.
In the contracting phase, radio halos are located at the hot region
of ICM, or between two substructures. 
In the expanding phase, on the other hand, radio halos are 
located between two ICM hot regions and shows rather diffuse
distribution. 
\end{abstract}

\keywords{galaxies: clusters: general --- hydrodynamics--- 
intergalactic medium --- particle acceleration ---
radiation mechanisms: nonthermal}

\section{INTRODUCTION}\label{s:intro}

Some clusters of galaxies have diffuse non-thermal synchrotron 
radio halos, which extend in a $\sim$ Mpc scale 
(e.g., Giovannini et al. 1993; 
R\"{o}ttgering et al. 1997; Deiss et al. 1997).  
This indicates that there exists a
relativistic electron population with energy of a few GeVs 
(if we assume the magnetic field strength is an order of $\mu$G)
in intracluster
space in addition to the thermal intracluster medium (ICM). 
Furthermore, it is well known that such clusters of galaxies 
have evidences of recent major merger in
X-ray observations (e.g., Henriksen \& Markevitch 1996; Honda et al. 1996;
Markevitch, Sarazin, \& Vikhlinin 1999; Watanabe et al. 1999). 
In such clusters of galaxies with radio halos, 
non-thermal X-ray radiation due to 
inverse Compton (IC) scattering of cosmic microwave background (CMB) photons
by the same electron population is expected (Rephaeli 1979).
Indeed, non-thermal X-ray radiation was recently detected 
in a few rich clusters (e.g., Fusco-Femiano et al. 1999; 
Rephaeli, Gruber, \& Blanco 1999; Kaastra et al. 1999) 
and several galaxy groups (Fukazawa 1999)
although their origins are still controversial.
In addition to such relatively high energy non-thermal emission, 
diffuse extreme ultraviolet (EUV) emission is detected from a number of
clusters of galaxies  (Lieu et al. 1996; Mittaz, Lieu, \& Lockman 1998; 
Lieu, Bonamente, \& Mittaz 1999).
Although their origins are also unclear, one hypothesis is that EUV emission
is due to IC emission of CMB.
If the hypothesis is right, this indicates existence of relativistic electrons 
with energy of several handled MeVs in intracluster space.

The origin of such  relativistic electrons is still
unclear. Certainly, there are point sources like radio galaxies in clusters
of galaxies, which produces such a electron population. However, the electrons 
can spread only in a few kpc scale by diffusion during the IC cooling time.
Clearly, this cannot explain typical spatial
size of radio halos. One possible solution of this problem is a secondary
electron model, where the electrons are produced through
decay of charged pions induced by the
interaction between relativistic protons and thermal protons in ICM
(Dennison 1980). In the secondary electron model, however, 
too much gamma-ray emission 
is produced through neutral pion decay to fit the Coma cluster results 
(Blasi \& Colafrancesco 1999). 
Moreover, the secondary electron model cannot explicitly explain 
the association between merger and radio halos.

From N-body + hydrodynamical simulations, it is expected that there exist 
shock waves and strong bulk-flow motion in ICM during merger 
(e.g, Schindler \& M\"{u}ller 1993; Ishizaka \& Mineshige 1996;
Takizawa 1999, 2000). 
This suggests that relativistic electrons are produced
at around the shock fronts through 1st order Fermi acceleration and 
that propagation of the shock waves and bulk-flow of ICM are responsible 
for extension of radio halos.  
Obviously, the merger shock acceleration model can explicitly
explain the association between merger and radio halos.
However, such hydrodynamical effects on time evolution
and spatial distribution of relativistic electrons during merger
are not properly considered in previous studies.

In this paper, we investigate the evolution of a relativistic electron 
population and non-thermal emission in the framework of the merger shock
acceleration model.
We perform N-body + hydrodynamical simulations, explicitly considering 
the evolution of a relativistic electron population produced at 
around the shock fronts.

The rest of this paper is organized as follows. In \S \ref{s:order} we 
show order estimation about the spatial size of radio halos.
In \S \ref{s:models} we describe the adopted numerical methods and initial
conditions for our simulations. In \S \ref{s:results} we present the results.
In \S \ref{s:conclusions} we summarize the results and discuss their 
implications.

\section{ORDER ESTIMATION}\label{s:order}

In this section, we estimate several kinds of spatial length scale 
relevant to the extent of cluster radio halos.

\subsection{Diffusion Length}

According to Bohm diffusion approximation, a diffusion coefficient is,
\begin{eqnarray}
   \kappa = \frac{ \eta E_{\rm e} c }{3 e B}, \label{eq:kappa}
\end{eqnarray}
where, $\eta$ is an enhanced factor from Bohm diffusion limit,
$E_{\rm e}$ is the total energy of an electron,
$c$ is the velocity of light, $e$ is the electron charge,
and $B$ is the magnetic field strength 
of intracluster space. Since IC scattering of CMB photons is the dominant 
cooling process for electrons with energy of $\sim$ GeVs in typical
intracluster conditions (Sarazin 1999), electron cooling 
time in this energy range is,
\begin{eqnarray}
  t_{\rm IC} = 1.1 \times 10^9 {\rm yr} \biggr( \frac{E_{\rm e}}{\rm GeV} 
                                         \biggl)^{-1},
  \label{eq:tic}
\end{eqnarray}
where we assume that the cluster redsift is much less than unity.
Thus, the diffusion length within the cooling time is
\begin{eqnarray}
  L_{\rm diff} \sim \sqrt{ \kappa t_{\rm IC} }
               \sim 1.1 \times 10^{-4} {\rm Mpc} 
               \biggr( \frac{\eta}{10^2} \biggl)^{1/2} 
               \biggr( \frac{B}{\mu {\rm G}} \biggl)^{-1/2}. \label{eq:ldiff}
\end{eqnarray}
This is much less than the spatial size of radio halos. From this reason,
electrons which are leaking from point sources such like AGNs cannot be 
responsible for the radio halos.

\subsection{Shock Wave Propagation Length}

From N-body + hydrodynamical simulations of cluster mergers, 
the propagation speed of shock wave is an order of 
$\sim 1000$  km s$^{-1}$ (Takizawa 1999). 
Thus, the length scale where the shock front 
propagates during the cooling time of equation (\ref{eq:tic}) is,
\begin{eqnarray}
	L_{\rm shock.prop} \sim 1.1 {\rm Mpc} 
                      \biggr( 
                      \frac{ v_{\rm shock} }{1000 {\rm km /s}}
                      \biggl)
		      \biggr( 
                      \frac{ E_{\rm e} }{ {\rm GeV} }
                      \biggl)^{-1},
\end{eqnarray}
where $v_{\rm shock}$ is the propagation speed of the shock front.
Roughly speaking, $L_{\rm shock.prop}$ is related to the extent of 
radio halos along to the collision axis since the accelerated electrons 
can emit synchrotron radio radiation only during the cooling time behind
the shock front.

\subsection{Spatial Size of Shock Surfaces}

In cluster merger, the shock front spread over in a cluster scale 
($\sim$ Mpc). Thus, even if the shock is nearly standing, radio halos
spread in a Mpc scale. Roughly speaking, this is related to the extent of
radio halos perpendicular to the collision axis. 

In the merger shock acceleration model, therefore, propagation of shock 
fronts and spatial size of shock surfaces play more crucial role to 
the extent of radio halos than diffusion. Furthermore, the model naturally 
produce 3-dimensionally extended radio halos in a Mpc scale.

\section{MODELS}\label{s:models}

We consider the merger of two equal mass ($0.5 \times 10^{15} M_{\odot}$)
subclusters. In order to calculate the evolution of ICM, 
we use the smoothed-particle hydrodynamics (SPH) method. 
Each subcluster is represented by 5000 N-body particles and 5000 SPH
particles. 
The initial conditions for ICM and N-body components are the
same as those of Run A in Takizawa (1999). The numerical methods and 
initial conditions for N-body and hydrodynamical parts are fully
described in \S 3 of Takizawa (1999). Our code is fully 3-dimensional.

To follow the evolution of a relativistic electron population, we should solve 
the diffusion-loss equation (see Longair 1994) for each SPH particle.
Since the diffusion term is negligible as mentioned in \S \ref{s:order},
the equation is,
\begin{eqnarray}
	\frac{d N(E_{\rm e}, t)}{dt} = \frac{\partial}{\partial E_{\rm e}}
        [b(E_{\rm e}, t)N(E_{\rm e}, t)] + Q(E_{\rm e}, t),
         \label{eq:diffloss}
\end{eqnarray}
where $N(E_{\rm e}, t)dE_{\rm e}$ is the total number of relativistic 
electrons per a SPH particle with kinetic energies in the range $E_{\rm e}$ 
to $E_{\rm e}+dE_{\rm e}$ (hereafter, we denote kinetic energy of an electron 
to $E_{\rm e}$), $b(E_{\rm e}, t)$ is 
the rate of energy loss for a single electron with an energy of $E_{\rm e}$, 
and $Q(E_{\rm e}, t)dE_{\rm e}$ gives the rate of production of new 
relativistic electrons per a SPH particle. 

According to the standard theory of 1st order Fermi acceleration, we assume
that $Q(E_{\rm e}, t) \propto E_{\rm e}^{- \alpha}$, where $\alpha$ is 
described as $(r+2)/(r-1)$ using the compression ratio of the shock front, 
$r$. For the shocks appeared in this simulation, the ratio is roughly 
$\sqrt{10}$ (Takizawa 1999),
which provides $\alpha =2.4$. Since it is very difficult to monitor 
the compression ratio at the shock front for each SPH particle 
per each time step, we neglect the time dependence of $\alpha$.
The influence of the changes in $\alpha$ on the results is discussed 
in \S \ref{s:conclusions}.
The normalization of $Q(E_{\rm e}, t)$ is proportional to the
artificial viscous heating, which is nearly equal to the shock heating.
We generate the relativistic electrons everywhere even if explicit 
shock structures do not appear in the simulation.
We assume that sub-shock exists where a fluid  element has 
enough viscous heating. Such sub-shocks are recognized in higher 
resolution simulations (e.g. Roettiger, Burns, \& Stone 1999).
We assume that total kinetic energy of accelerated electrons
from $E_{\rm e} = 0$ to $+\infty$ is 5\% of the viscous energy, 
which is consistent with the recent TeV
gamma ray observational results for the galactic supernova remnant 
SN 1006 (Tanimori et al. 1998; Naito et al. 1999). 
Note that equation \ref{eq:diffloss} for the evolution of a relativistic
electron population is linear in $N( E_{\rm e}, t)$. Thus, it is easy to 
rescale our results of $N(E_{\rm e}, t)$ if we choose other parameters 
for the acceleration efficiency. We neglect energy loss of thermal ICM 
due to the acceleration.

For $b(E_{\rm e}, t)$, we consider IC scattering of the CMB photons, 
synchrotron losses, and Coulomb losses.
We neglect bremsstrahlung losses for simplicity, which is a good 
approximation in typical intracluster conditions (Sarazin 1999).
Then, if we ignore weak energy dependence of coulomb losses, 
the loss function $b(E_{\rm e}, t)$ becomes 
\begin{eqnarray}
	b(E_{\rm e}, t) = b_c(t) + b_1(t) E_{\rm e}^2, 
\end{eqnarray}
where $b_c(t) = 7.0 \times 10^{-16} (n_{\rm e}(t) / {\rm cm}^{-3})$
and $b_1 = 2.7 \times 10^{-17}+2.6 \times 10^{-18} (B(t)/\mu {\rm G})^2$
(if $b(E_{\rm e}, t)$ and $E_{\rm e}$ are given in units of GeV s$^{-1}$ and
GeV, respectively.).
In the above expressions, $n_{\rm e}$ is the number density of ICM electrons
and $B$ is strength of the magnetic field.

To integrate equation (\ref{eq:diffloss}) with the Courant and viscous
timestep control (see Monaghan 1992), we use the analytic solution as follows.
First, we integrate equation (\ref{eq:diffloss}) from $t$ to $t+\Delta t$, 
regarding the second term on the right-hand side as being negligible small. 
Then,
\begin{eqnarray}
      N(E_{\rm e}, t + \Delta t) = \left\{ 
                       \begin{array}{@{\,}ll}
      N( E_{{\rm e},0}, t) \frac{b(E_{{\rm e},0})}{b(E_{\rm e})}, 
                           & \mbox{ ($E_{\rm e}<E_{{\rm e}, {\rm max}}$), } \\
      0,                   
                           & \mbox{ ($E_{\rm e}>E_{{\rm e}, {\rm max}}$), }
                       \end{array}
                           \right.  
\end{eqnarray}
where 
\begin{eqnarray}
    E_{{\rm e},0} &=& \sqrt{ \frac{b_c}{b_1} } \tan \biggr( 
    \arctan{ \sqrt{ \frac{b_1}{b_c}} E_{\rm e} } + \sqrt{b_c b_1} \Delta t
                                        \biggl),\\
    E_{{\rm e}, {\rm max}} &=& \sqrt{\frac{b_c}{b_1}} 
         \frac{1}{ \tan {\sqrt{b_1 b_c} \Delta t}}.
\end{eqnarray}
Next, we add the contribution from the second term to the above 
$N(E_{\rm e}, t + \Delta t)$ using the second-order Runge-Kutta method.
In the present simulation, $N(E_{\rm e}, t)$ is calculated on logarithmically 
equally spaced 300 points in the range $E_{\rm e} = 0.05$ to $50$ GeVs
for each SPH particle.

Magnetic field evolution is included by means of the following method.
We assume initial magnetic pressure is 0.1 \% of ICM thermal pressure. 
This corresponds to $B = 0.1 \mu {\rm G}$ in volume-averaged magnetic field 
strength. For Lagrangean evolution of $B$, due to the frozen-in assumption
we apply $B(t)/B(t_0) =  ( \rho_{\rm ICM}(t)/\rho_{\rm ICM}(t_0) )^{2/3}$.
Field changes due to the passage of the shock waves is not
considered in this paper. The change may depend on field configuration at 
the shock front and have value of $\sim 1-4$. However, it is difficult
to examine it in the present simulation even under high $\beta$ condition.
We will try this problem in the future paper.

Our model implies continuous production of power law distributed 
relativistic electrons at around the shock fronts.
This is valid only when $\Delta t_{\rm acc}$ is sufficiently
shorter than the dynamical timescale of the system ($\sim 10^9$ yr),
where $\Delta t_{\rm acc}$ denotes acceleration time 
in which $Q(E_{\rm e},t)$ becomes power law distribution.
It is presented in the framework of the standard shock acceleration theory as
$\Delta t_{\rm acc}=3 r u^{-2} (r-1)^{-1} (\kappa_{1}+r \kappa_{2})$
, where $u$ is the flow velocity of the upstream of the shock front, 
and $\kappa_{1, 2}$ are diffusion coefficients 
of the upstream and downstream, respectively (see e.g. Drury 1983).
Assuming $B_{1}=B_{2}$ and Bohm diffusion approximation 
of equation (\ref{eq:kappa}),
\begin{eqnarray}
   \Delta t_{\rm acc}=1.9 \times 10^{2} {\rm yr} 
        \biggr( \frac{E_{\rm e}}{\rm GeV} \biggl) 
        \biggr( \frac{\eta}{10^{2}} \biggl) 
        \biggr( \frac{u}{10^3 {\rm km \ s}^{-1}} \biggl)^{-2} 
        \biggr( \frac{B}{\rm \mu G} \biggl)^{-1}.
\end{eqnarray}
This value is certainly much shorter than the dynamical timescale.

\section{RESULTS}\label{s:results}

\begin{figure}
	\epsfysize = 14.0 cm
        \centerline{\epsfbox{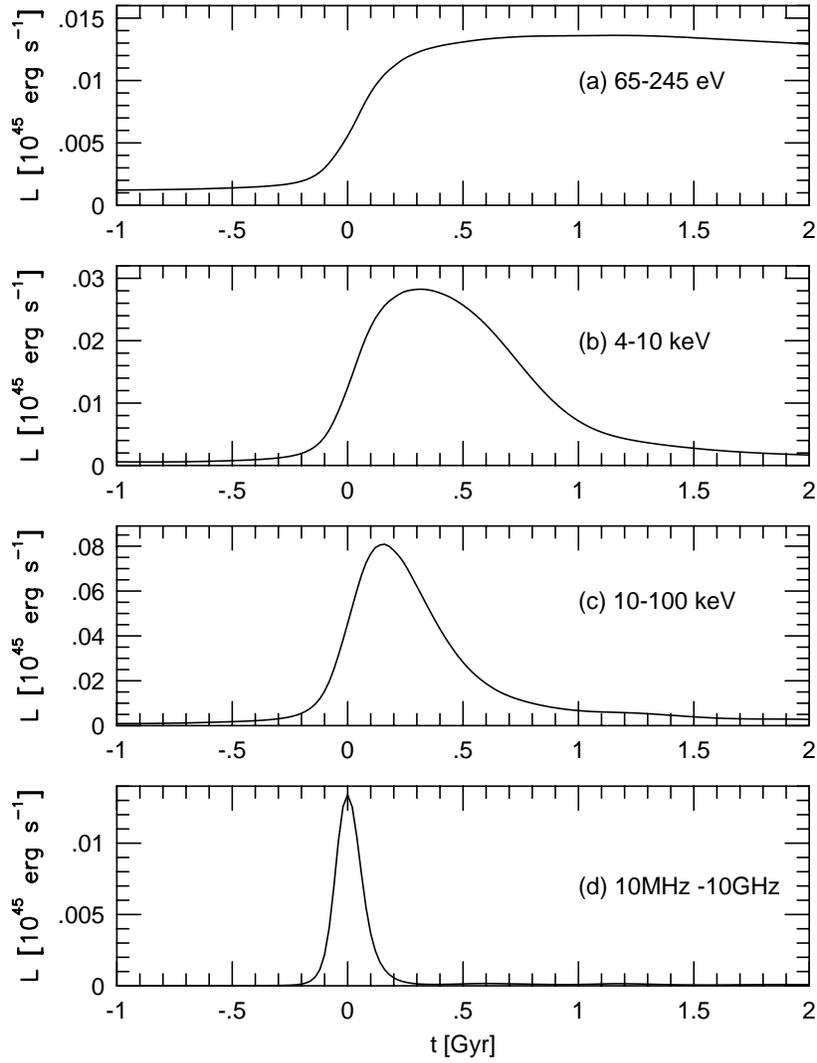}}
  \caption{The time evolution of non-thermal emission for various
            energy band: from top to bottom, inverse Compton scattering 
            of EUVE band (65-245 eV), soft X-ray band (4-10 keV), and 
            hard X-ray band (10-100 keV), and synchrotron radio emission 
            (10 MHz - 10 GHz). The times are relative to the most 
            contracting epoch.}
  \label{fig:nthe}
\end{figure}

Figure \ref{fig:nthe} shows the time evolution of non-thermal 
emission for various
energy band: from top to bottom, IC emission of the Extreme Ultraviolet
Explorer (EUVE) band (65-245 eV),
soft X-ray band (4-10 keV), and hard X-ray band (10-100 keV), and 
synchrotron radio emission (10 MHz - 10 GHz). The times are relative to the
most contracting epoch. The calculation of the luminosity for each band
is performed in the simplified assumption that electrons radiate at a 
monochromatic
energy given by $E_{\rm X} = 2.5 {\rm keV} (E_{\rm e}/{\rm GeV})^2$ and 
$\nu = 3.7 {\rm MHz} (B/\mu {\rm G}) (E_{\rm e}/{\rm GeV})^2$
for IC scattering and synchrotron emission, respectively.
Since cooling time is roughly proportional to $E_{\rm e}^{-1}$ 
in these energy range, the higher the radiation energy of IC emission is,
the shorter duration of luminosity increase is. In other words, luminosity
maximum comes later for lower energy band. 
Hard X-ray and radio emissions come from the electrons with almost
the same energy range. The luminosity maximum in the hard X-ray band, 
however, comes slightly after the most contracting epoch.
On the other hand, radio emission becomes maximum at most contracting 
epoch since the change of magnetic field due to the compression
and expression plays an more crucial role than the increase of relativistic
electrons. In any cases, radio halos and hard X-ray are well associated
to merger phenomena. They are observable only when thermal ICM have 
definite signatures of mergers such as complex temperature structures,
non-spherical and elongated morphology, or substructures.
Soft X-ray emission, which is observable only in clusters (or groups)
with relatively low temperature ($\simeq 1$ keV) ICM, is still luminous
in $\sim$ 1 Gyr after the merger. Thus, the association of mergers in this 
band is weaker than in the hard X-ray band.
Moreover, EUV emission 
continues to be luminous after the signatures of the merger have been 
disappeared in the thermal ICM.

\begin{figure}
	\epsfysize = 10.0 cm
        \centerline{\epsfbox{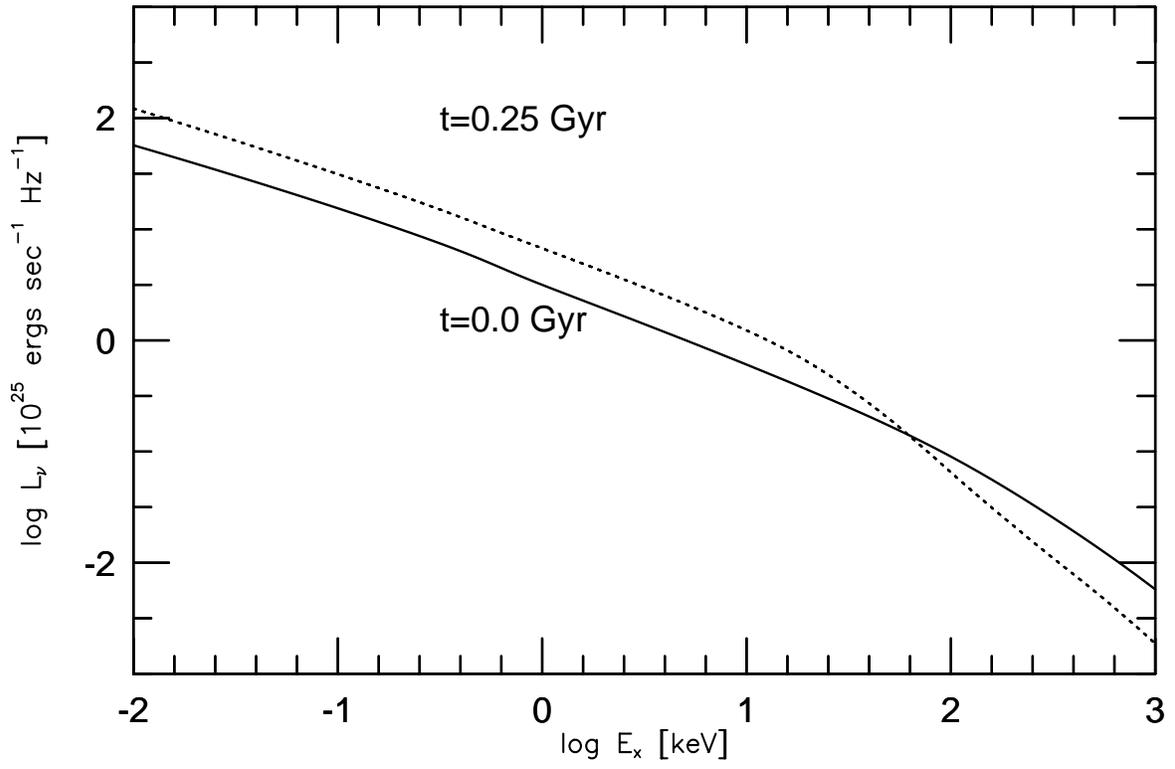}}
   \caption{The inverse Compton scattering spectra at $t=0.0$ (solid lines) 
            and $0.25$ (dotted lines).}
   \label{fig:icspec}
\end{figure}

Figure \ref{fig:icspec} shows the IC spectra at $t=0.0$ 
(solid lines) and $0.25$ (dotted lines). In lower energies, 
$L_{\nu} \propto \nu^{-0.7}$, which is originated from the electron
source spectrum $Q(E_{\rm e}) \propto E_{\rm e}^{-2.4}$. 
On the other hand, in higher
energies, the spectrums become close to steady solution, 
$L_{\nu} \propto \nu^{-1.2}$, owing to the IC and synchrotron 
losses (Longair 1994). The break point of the spectrum moves toward 
lower energies as time proceeds.

\begin{figure}
	\epsfysize = 6.0 cm
        \centerline{\epsfbox{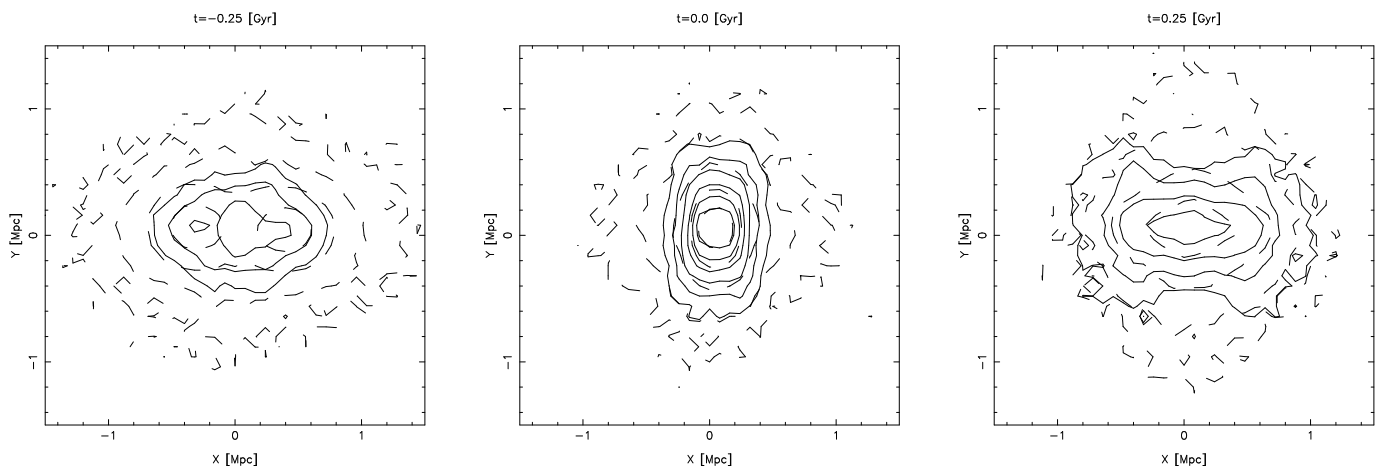}}
   \caption{Snapshots of synchrotron radio (10MHz-10GHz) surface brightness 
            distribution (solid contours) and X-ray one of thermal ICM 
            (dashed contours) seen from the direction perpendicular to the 
            collision axis. Contours are equally spaced on a logarithmic 
            scale and separated by a factor of 7.4 and 20.1 for radio and 
            X-ray maps, respectively.}
   \label{fig:syncmap}
\end{figure}

Figure \ref{fig:syncmap} shows the snapshots of synchrotron radio 
(10MHz-10GHz) surface brightness distribution (solid contours)
and X-ray one of thermal ICM (dashed contours)
seen from the direction perpendicular to the collision axis. 
Contours are equally spaced on a logarithmic scale and separated 
by a factor of 7.4 and 20.1 for radio and X-ray maps, respectively. 
At $t=-0.25$, the main shocks are located between the two X-ray peaks
and relativistic electrons are abundant there. Thus, the radio emission 
peak is located between the two X-ray peaks although the magnetic field 
strength there is weaker. 
At $t= 0.0$, relativistic electrons are
concentrated around the central region since the main shocks
are nearly standing
and located near $X \simeq \pm 0.2$. Furthermore, gas infall compress ICM and 
the magnetic field. Thus, radio distribution shows rather strong 
concentration. In these phase (at $t=-0.25$ and $0.0$), 
the radio halo is located at the high
temperature region of ICM. On the other hand, at $t=0.25$, relativistic
electron distribution becomes rather diffuse since fresh relativistic 
electrons are producing in the outer regions as the shock waves propagate 
outwards. At $t=0.25$ the main shocks are located at $X \simeq \pm 1$. 
Between the shock fronts rather diffuse radio emission is seen. 
In this phase, the radio halo is located between two high temperature 
regions of ICM.

\begin{figure}
	\epsfysize = 6.0 cm
        \centerline{\epsfbox{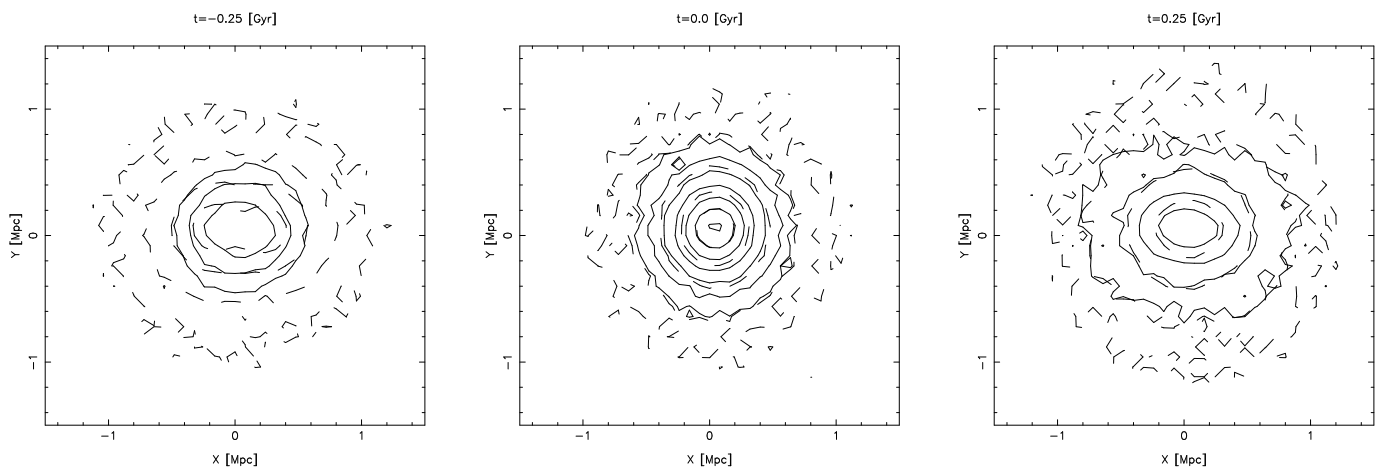}}
   \caption{Same as figure \ref{fig:syncmap}, but for seen from the direction
            tilted at an angle of $30^{\circ}$ with respect to the collision
            axis.}
   \label{fig:syncmap2}
\end{figure}

As described above, the morphology of the radio halo is strongly 
depending on the phase of the merger when viewed from the direction 
perpendicular to the collision axis. When viewed nearly along 
the collision axis, however, this is not the case. Figure \ref{fig:syncmap2} 
shows the same as figure \ref{fig:syncmap}, but for seen from the direction 
tilted at an angle of $30^{\circ}$ with respect to the collision axis. 
Radio and X-ray morphology are similar each other in all phases. 
When the cluster is viewed along the collision axis, the distribution 
of relativistic electrons roughly follows that of the thermal ICM since 
the shock fronts face to the observers and spread over the cluster. 
The distribution of magnetic field strength also roughly follows that of 
the thermal ICM. Therefor, the radio morphology follows X-ray one.

\begin{figure}
	\epsfysize = 10.0 cm
        \centerline{\epsfbox{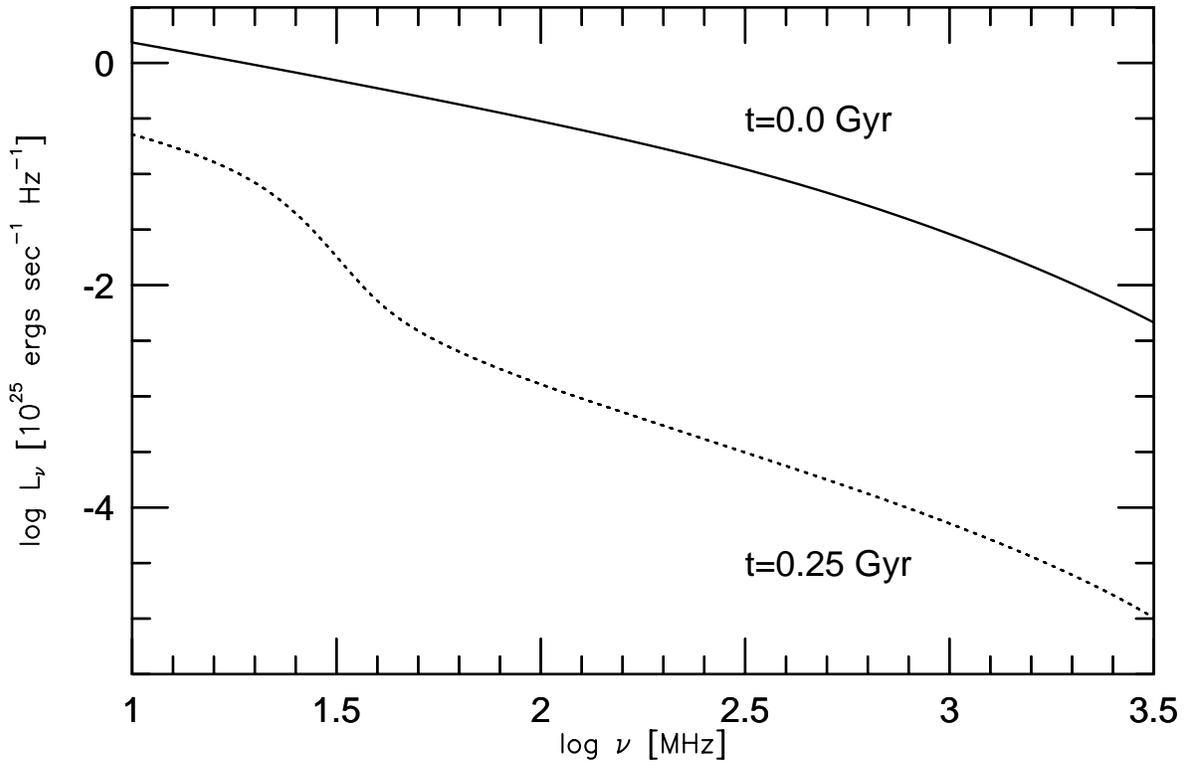}}
   \caption{The synchrotron radiation spectra at $t=0.0$ (solid lines) and 
            $0.25$ (dotted lines).}
   \label{fig:syncspec}
\end{figure}

Figure \ref{fig:syncspec} shows the synchrotron radiation spectra
at $t=0.0$ (solid lines) and $0.25$ (dotted lines). At $t=0.0$, the 
synchrotron spectrum follows that of IC emission.
On the other hand, at $t=0.25$ there exits a bump at lower energies 
in the spectrum, which cannot be seen in that of IC emission.
Emissivity of synchrotron radiation depends on not only relativistic 
electron density but also magnetic energy density, 
which is larger in the central region in this model.
On the other hand, IC emissivity depends on the electron density and
CMB energy density, which is homogeneous.
Thus, the total synchrotron spectrum is more like that in the 
central region than the total IC spectrum. At $t=0.0$, since relativistic 
electrons are centered, the emission from the outer region is
negligible for both spectra. Thus, similar results are given.
On the other hand, at $t=0.25$, propagation of shocks makes a diffuse
distribution of relativistic electrons. 
Therefor, the contribution from the outer region is not negligible
for the total IC spectrum while the total synchrotron spectrum
is still biased the central region as seen in figure \ref{fig:syncmap}.
In the central region, however, electrons produced by the main shocks
at $t \simeq 0$ with energies more than $\sim$ several GeVs
have already cooled down. Thus the spectrum in the central region
have a bump in lower energies, which is present in the total spectrum.
The emission above $\sim 30$ MHz is mainly due to the electrons
produced by the sub-shocks there. If such sub-shocks do not exist,
the emission near the main shocks, where the magnetic field strength 
is rather weaker, can be seen in this energy range.
Note that this feature of the synchrotron spectrum is sensitive
to the spatial distribution of magnetic field.

\section{CONCLUSIONS AND DISCUSSION}\label{s:conclusions}

We have investigated evolution of non-thermal emission from relativistic 
electrons accelerated at around the shock fronts during merger of clusters of 
galaxies. Hard X-ray and radio radiations are luminous only while
merger signatures are left in thermal ICM. 
Hard X-ray radiation becomes maximum in the later stage of merger.
In our simulation, radio emission is the most luminous at the most contracting 
epoch. This is due to the magnetic field amplification by compression.
According to the recent magnetohydrodynamical simulations
(Roettiger, Stone, \& Burns 1999), however, 
it is possible that the field amplification occurs as the bulk
flow is replaced by turbulent motion in the later stages of merger.
If this is effective in real clusters, radio emission can increase
by a factor of two or three
than our results in the later stages of merger.
EUV emission is still luminous after the merger signatures 
have been disappeared
in thermal ICM. This is consistent with the EUVE results.

Morphological relation between radio halos and ICM hot regions is described
as follows. In the contracting phase, radio halos are located at the hot 
regions of thermal ICM, or between two substructures 
(see the left panel of figure \ref{fig:syncmap}). This may correspond 
to A2256 (R\"{o}ttgering et al. 1994). In the expanding phase, on the other 
hand, radio halos are located between the two hot regions of ICM and show 
rather diffuse distribution (see the right panel of figure \ref{fig:syncmap}). 
This may correspond to Coma (Giovannini et al. 1993; Deiss et al. 1997) and 
A2319 (Feretti, Giovannini, \& B\"{o}hringer 1997). In the further later phase,
the shock fronts reach outer regions and the GeV electrons are already cooled 
in the central parts. Then, radio halos are located in the cluster 
outer regions near the shock fronts and we cannot detect radio emission
in the central part of the cluster. However, observational correlation 
between ICM hot regions and radio halos is not clear since the electron 
temperature there is significantly lower than the plasma mean temperature 
due to the relatively long relaxation time between ions and electrons
(Takizawa 1999, 2000). Note that until now we could only find the electron
temperature through X-ray observations. This may correspond to A3667 
(R\"{o}ttgering et al. 1997). It is possible that such radio {\lq}halos{\rq} 
located in the outer regions are classified into radio {\lq}relics{\rq} 
since their radio powers and spatial scales becomes weaker and smaller 
than those of typical radio halos, respectively. When the cluster is 
viewed nearly along the collision axis, however, such morphological 
relations between radio halo and ICM are unclear and radio and 
X-ray distributions become similar each other.

We neglect the changes of the spectral index in the electron source term.
Since the mach number is gradually increasing as merger proceeds 
(Takizawa 1999), the spectrum of relativistic electrons becomes 
flatter as time proceeds. We believe that such changes in the spectral 
index does not influence 
our results seriously because most of relativistic electrons are 
produced in the central high density region, where the mach number 
is almost constant. In the later stage of merger, however,
contribution of relativistic electrons produced in the outer region cannot
be negligible in higher energy range ($\sim 10$ GeV) since cooling time 
is relatively short. Thus, it is probable that the inverse Compton spectrum
in the hard X-ray ($\sim 10-100$ keV) becomes flatter in the later stages. 

The lower energy part of the electron spectrum can emit 
hard X-ray through bremsstrahlung. Whether IC scattering or bremsstrahlung 
is dominant in the hard X-ray range is depending on the shape of
the electron spectrum. Roughly speaking, when the spectrum of relativistic 
electrons is flatter than $E_{\rm e}^{-2.5}$, 
IC scattering dominates the other 
and vice versa (see Appendix). Furthermore, the shape of the electron 
spectrum in the lower energy part is flatter than the originally injected
form since the cooling time due to the coulomb loss is proportional to
$E_{\rm e}$ and very short (Sarazin 1999). In the present simulation, 
therefor, it is most likely that the contribution 
of the bremsstrahlung components in the hard X-ray range is negligible. 
More detailed calculations, including nonlinear effects for the shock 
acceleration (Jones \& Ellison 1991), by Sarazin \& Kempner (1999)
show that IC scattering is dominant in the hard X-ray when the accelerated 
electron momentum spectrum is flatter than 
$p_{\rm e}^{-2.7}$, which corresponds to the electron energy spectrum of 
$E_{\rm e}^{-2.7}$ in the fully relativistic range. Thus, the bremsstrahlung 
contribution in the hard X-ray should be considered in mergers 
with low mach numbers.

A merger shock acceleration model also predicts some gamma-ray emission.
Electrons which radiate EUV due to IC scattering also emit 
$\sim 100$ MeV gamma-ray through bremsstrahlung. Furthermore, it is most
likely that protons as well as electrons are accelerated at around
the shock fronts. Such high energy protons also produce gamma-rays
peaked at $\sim 100$ MeV through decay of neutral pions. 
Although the energy density ratio between electrons and protons
in acceleration site is uncertain, the contributions of protons 
and the bremsstrahlung to the emission become 
important in the higher energies. We think that it is interesting
to investigate from hundreds MeV to multi TeV emissions, which are
observable with operating instruments such as {\it EGRET}, ground-based 
air \v{C}erenkov telescopes, and planning projects like {\it GLAST} 
satellites. However, since the diffusion length of protons within the 
cooling time is much longer than that of electrons, treatment of the 
diffusion-loss equation for protons are more complex than the model 
in this paper.

\acknowledgements
We would like to thank Drs. Y. Fukazawa and S. Shibata for helpful comments.
MT thanks Drs. S. Mineshige and T. Shigeyama for continuous encouragement.
MT is also grateful to S. Tsubaki for fruitful discussion.

\appendix
\section{ESTIMATION OF THE BREMSSTRAHLUNG CONTRIBUTION IN THE HARD X-RAY RANGE}

We estimate the contribution 
of the bremsstrahlung form the lower energy part of the electron spectrum 
to the hard X-ray emission in our model,
which is neglected in this paper.
Although the crude estimations discussed here are order-of magnitudes,
it is helpful to explain which mechanism, 
IC scattering and bremsstrahlung radiation, 
is dominant.

The standard theory of 1st order Fermi acceleration 
provides power law spectrum in momentum distribution.
We, therefore, assume the momentum spectrum of accelerated electrons 
has a form of 
\begin{eqnarray}
  \frac{dN_{\rm e}}{dP_{\rm e}}=
  c N_{\rm 0} \left(\frac{P_{\rm e}}{m_{\rm e} c}\right)^{- \alpha}\ \ 
  {\rm [cm^{-3}\ (eV\ {\mit c}^{-1})^{-1}]}\ ,
  \label{eq:npspec}
\end{eqnarray}
where $P_{\rm e}$ is an electron momentum, $c$ is the speed of light, 
and $m_{\rm e}$ is the electron rest mass.
Using the relation between the momentum $P_{\rm e}$ 
and the kinetic energy $E_{\rm e}$, 
$E_{\rm e} = (P_{\rm e}^2 c^2 + m_{\rm e}^2 c^4)^{1/2} - m_{\rm e} c^2$,
the spectrum for kinetic energy is given by 
\begin{eqnarray}
  \frac{dN_{\rm e}}{dE_{\rm e}}=
  N_{\rm 0} \frac{E_{\rm e} + m_{\rm e} c^2}{m_{\rm e} c^2} 
  \left(\frac{P_{\rm e}}{m_{\rm e} c}\right)^{- \alpha - 1}\ \ 
  {\rm [cm^{-3}\ (eV)^{-1}]}\ .
  \label{eq:nespecgen}
\end{eqnarray}
For relativistic electrons, $E_{\rm e} \gg m_{\rm e} c^2$, 
the spectrum becomes
\begin{eqnarray}
  \frac{dN_{\rm e}}{dE_{\rm e}}=
  N_{\rm 0} \left(\frac{E_{\rm e}}{m_{\rm e} c^2}\right)^{- \alpha}\ ,
  \label{eq:nespecrel}
\end{eqnarray}
which is consistent with an assumption of $Q(E_{\rm e},t)$
in section 3.
Since the hard X-ray photons are produced 
from CMB photon field 
via IC scattering of relativistic electrons in our model,
we use this form for the estimation of IC emissivity.
For non-relativistic electrons, $E_{\rm e} \ll m_{\rm e} c^2$, 
the spectrum becomes
\begin{eqnarray}
  \frac{dN_{\rm e}}{dE_{\rm e}}=
  2^{- \frac{\alpha + 1}{2}} N_{\rm 0}
  \frac{E_{\rm e} + m_{\rm e} c^2}{m_{\rm e} c^2} 
  \left(\frac{E_{\rm e}}{m_{\rm e} c^2}\right)^{- \frac{\alpha + 1}{2}}\ ,
  \label{eq:nespecnrel}
\end{eqnarray}
where we use the relation $E_{\rm e}=P_{\rm e}^2/(2 m_{\rm e})$.
Since the bremsstrahlung radiation at the hard X-ray range is emanated 
from electrons with almost the same energy range,
we use this form for the estimation of bremsstrahlung emissivity.

For simplicity, 
we approximate the emissivity, $\epsilon$, 
for both IC and bremsstrahlung processes to be,
\begin{eqnarray}
  \epsilon=
  {\frac{dN_{\rm e}}{dE_{\rm e}}}
  \frac{dE_{\rm e}}{d{\varepsilon}_{\rm \gamma}}
  {\left| \frac{dE_{\rm e}}{dt} \right|}\ \ 
  {\rm [erg\ s^{-1}\ cm^{-3}\ eV^{-1}]}\ ,
  \label{eq:emiss}
\end{eqnarray}
where the emission rate is assumed to be equal to the electron energy 
loss rate and $\varepsilon_{\gamma}$ is the photon energy.

To obtain the bremsstrahlung emissivity, 
we assume that an electron with energy $E_{\rm e}$
loses its energy to emit photon of energy 
$\varepsilon_{\rm \gamma}=E_{\rm e}$
after it has traversed one mean free path $X_{\rm 0}$.
Hence,
\begin{eqnarray}
  \frac{dE_{\rm e}}{d{\varepsilon}_{\rm \gamma}}=1,
  \label{eq:dedphotonbrem}
\end{eqnarray}
and
\begin{eqnarray}
  \left| \frac{dE_{\rm e}}{dt} \right| \sim
  \varepsilon_{\rm \gamma} \frac{v_{\rm e}}{X_{\rm 0}} 
  \ \ {\rm [erg\ s^{-1}]}\ ,
\end{eqnarray}
where $v_{\rm e}$ is electron velocity.
From $\sigma_{\rm T} n_{\rm 0} X_{\rm 0} \sim 1$, we approximate
\begin{eqnarray}
  \left| \frac{dE_{\rm e}}{dt} \right| \sim
  \varepsilon_{\rm \gamma} v_{\rm e} \sigma_{\rm T} n_{\rm 0}\ ,
  \label{eq:coolbrem}
\end{eqnarray}
where $\sigma_{\rm T}$ denotes the cross section of Thomson scattering 
and $n_{\rm 0}$ denotes the density of ambient matter.
Using equations (\ref{eq:nespecnrel}), (\ref{eq:emiss}), 
(\ref{eq:dedphotonbrem}), and (\ref{eq:coolbrem}),
we estimate the bremsstrahlung emissivity 
in the hard X-ray energy range, 
$\varepsilon_{\rm \gamma}=\varepsilon_{\rm HXR}$, as
\begin{eqnarray}
  \epsilon_{\rm brem} \sim
  2^{- \frac{\alpha}{2}} N_{\rm 0} 
  \sigma_{\rm T} c n_{\rm 0} m_{\rm e} c^2 
  \left(1+\frac{\varepsilon_{\rm HXR}}{m_{\rm e} c^2} \right)
  \left(\frac{\varepsilon_{\rm HXR}}{m_{\rm e} c^2} \right)
  ^{- \frac{\alpha}{2} + 1}\ ,
  \label{eq:emissbrem}
\end{eqnarray}

For IC scattering of CMB photons, 
the photon energy after the scattering 
by an electron with energy 
$E_{\rm e}=m_{\rm e} \gamma_{\rm e} c^2$
is approximated by single energy of 
$\varepsilon_{\rm \gamma}
={\gamma_{\rm e}}^2 \bar{\varepsilon}_{\rm CMB}$,
where $\bar{\varepsilon}_{\rm CMB}$ is peak energy of CMB spectrum.
Thus,
\begin{eqnarray}
  \frac{dE_{\rm e}}{d{\varepsilon}_{\rm \gamma}}
  =\frac{1}{2}
  \left(\frac{\bar{\varepsilon}_{\rm CMB}}{m_{\rm e} c^2} \right)
  ^{- \frac{1}{2}}
  \left(\frac{\varepsilon_{\rm \gamma}}{m_{\rm e} c^2} \right)
  ^{- \frac{1}{2}}\ .
  \label{eq:dedphotoic}
\end{eqnarray}
Since the scattering is in Thomson energy range 
($\bar{\varepsilon}_{\rm CMB} \gamma_{\rm e} \gg m_{\rm e} c^2$),
we set
\begin{eqnarray}
  \left| \frac{dE_{\rm e}}{dt} \right|
  =\frac{4}{3} \sigma_{\rm T} c {\gamma_{\rm e}}^2 
  \bar{\varepsilon}_{\rm CMB} n_{\rm CMB}
  \ \ {\rm [erg\ s^{-1}]}\ ,
  \label{eq:coolic}
\end{eqnarray}
where $n_{\rm CMB}$ is photon number density of CMB fields.
Using equations (\ref{eq:nespecrel}), (\ref{eq:emiss}), 
(\ref{eq:dedphotoic}), and (\ref{eq:coolic}),
we estimate the IC emissivity 
in the hard X-ray energy range, 
$\varepsilon_{\rm \gamma}=\varepsilon_{\rm HXR}$, as
\begin{eqnarray}
  \epsilon_{\rm IC} 
  =\frac{2}{3} N_{\rm 0} \sigma_{\rm T} c 
  \bar{\varepsilon}_{\rm CMB} n_{\rm CMB}
  \left(\frac{\bar{\varepsilon}_{\rm CMB}}{m_{\rm e} c^2} \right)
  ^{\frac{\alpha-3}{2}}
  \left(\frac{\varepsilon_{\rm HXR}}{m_{\rm e} c^2} \right)
  ^{- \frac{\alpha-1}{2}}\ .
  \label{eq:emissic}
\end{eqnarray}

From equations (\ref{eq:emissbrem}) and (\ref{eq:emissic}), 
we derive the emissivity ratio as
\begin{eqnarray}
  \frac{\epsilon_{\rm brem}}{\epsilon_{\rm IC}}
  =\frac{3}{2} \frac{n_{\rm 0}}{n_{\rm CMB}} 
  \left(1+\frac{\varepsilon_{\rm HXR}}{m_{\rm e} c^2} \right)
  \left(\frac{\bar{\varepsilon}_{\rm CMB}}{m_{\rm e} c^2} \right)
  ^{\frac{1}{2}}
  \left(\frac{\varepsilon_{\rm HXR}}{m_{\rm e} c^2} \right)
  ^{\frac{1}{2}}
  \left(2 \frac{\bar{\varepsilon}_{\rm CMB}}{m_{\rm e} c^2} \right)
  ^{- \frac{\alpha}{2}}
  \ .
  \label{eq:rate}
\end{eqnarray}
For $\epsilon_{\rm brem} < \epsilon_{\rm IC}$, 
we get the relation
\begin{eqnarray}
  \alpha < 
  \frac{
  2 \ln \left[ 
  \frac{3}{2} \frac{n_{\rm 0}}{n_{\rm CMB}} 
  \left(1+\frac{\varepsilon_{\rm HXR}}{m_{\rm e} c^2} \right)
  \frac{
  (\bar{\varepsilon}_{\rm CMB}\varepsilon_{\rm HXR})^{\frac{1}{2}}
  }{m_{\rm e} c^2}
  \right]
  }{
  \ln \left(
  \frac{2 \bar{\varepsilon}_{\rm CMB}}{m_{\rm e} c^2}
  \right)
  } \ .
  \label{eq:alpha}
\end{eqnarray}
Substituting typical values for ICM at $z \sim 0$,
$n_{\rm 0}=1 \times 10^{-3}\ \ {\rm cm^{-3}}$,
$n_{\rm CMB}=400\ \ {\rm cm^{-3}}$,
$\bar{\varepsilon}_{\rm CMB}=6.57 \times 10^{-4}\ \ {\rm eV}$,
and 
$\varepsilon_{\rm HXR}=10\ \ {\rm keV}$,
we obtain $\alpha < 2.5$.
At $\varepsilon_{\rm HXR}=100\ \ {\rm keV}$,
the spectrum of equation (\ref{eq:emissbrem}) becomes steeper 
as electrons enter the trans-relativistic energy range.
As a result, the bremsstrahlung emissivity is reduced
so that the larger index of electron spectrum is accepted 
for $\epsilon_{\rm brem} < \epsilon_{\rm IC}$.

\newpage

\clearpage

\end{document}